\documentclass{aa}
\usepackage{amssymb}
\usepackage{amsmath}
\usepackage{xspace}
\usepackage{graphicx}
\usepackage{enumerate}   
\usepackage{color}
\usepackage{hyperref}
\usepackage{multirow}

\usepackage{natbib}

\makeatletter
\newcommand*{\rom}[1]{\expandafter\@slowromancap\romannumeral #1@}
\makeatother

\newcommand{\n}{\nonumber}
\newcommand*\df{\mathop{}\!\mathrm{d}}

\usepackage{gensymb}

\usepackage{color}

\begin{document}

\title{Constraining the cosmic ray spectrum in the vicinity of the supernova remnant W28: from sub-GeV to multi-TeV energies}
\author{V. H. M. Phan$^1$, S. Gabici$^{1}$, G. Morlino$^{2}$, R. Terrier$^{1}$, J. Vink$^{3}$,  J. Krause$^{1}$, and M. Menu$^4$} 
\institute{$^{1}$APC, Universit\'e Paris Diderot, CNRS/IN2P3, CEA/Irfu, Observatoire de Paris, Sorbonne Paris Cit\'e, France \\
$^{2}$INAF/Osservatorio Astrofisico di Arcetri, L.go E. Fermi 5, 50125 Firenze, Italy\\
$^3$GRAPPA \& Anton Pannekoek Institute for Astronomy, University of Amsterdam, Science Park 904, 1098 XH Amsterdam, The Netherlands\\
$^4$Laboratoire de Physique des Plasmas, \'Ecole Polytechnique,
Univ. Paris-Sud, Universit\'e Paris-Saclay, F-91128 Palaiseau Cedex, France\\
}

\authorrunning{Phan et al.}
\titlerunning{Cosmic rays in the vicinity of W28}
\abstract
{Supernova remnants interacting with molecular clouds are ideal laboratories to study the acceleration of particles at shock waves and their transport and interactions in the surrounding interstellar medium.}
{In this paper, we focus on the supernova remnant W28, which over the years has been observed in all energy domains from radio waves to very-high-energy gamma rays. The bright gamma-ray emission detected from molecular clouds located in its vicinity revealed the presence of accelerated GeV and TeV particles in the region. An enhanced ionization rate has also been measured by means of millimetre observations, but such observations alone cannot tell us whether the enhancement is due to low energy (MeV) cosmic rays (either protons or electrons) or the X-ray photons emitted by the shocked gas.
The goal of this study is to determine the origin of the enhanced ionization rate and to infer from multiwavelength observations the spectrum of cosmic rays accelerated at the supernova remnant shock in the unprecedented range spanning from MeV to multi-TeV particle energies.
 }
{We developed a model to describe the transport of X-ray photons into the molecular cloud, and we fitted the radio,  millimeter, and gamma-ray data to derive the spectrum of the radiating particles.  }
{The contribution from X-ray photons to the enhanced ionization rate is negligible, and therefore the ionization must be due to cosmic rays.
Even though we cannot exclude a contribution to the ionization rate coming from cosmic ray electrons, we show that a scenario where cosmic ray protons explain both the gamma-ray flux and the enhanced ionization rate provides the most natural fit to multiwavelength data.
This strongly suggests that the intensity of CR protons is enhanced in the region for particle energies in a very broad range covering almost 6 orders of magnitude: from $\lesssim 100$ MeV up to several tens of TeV.
}
{}


\maketitle
\section{Introduction}

SuperNova Remnants (SNRs) interacting with Molecular Clouds (MCs) are ideal laboratories to study the acceleration of particles at astrophysical shocks \citep{gabici2015}. The study of such systems is of particular importance in connection with the problem of the origin of Galactic Cosmic Rays (CRs). This is because CRs are believed to be accelerated at SNR shocks, and injected in the interstellar medium with an energy spectrum which is a power law in momentum over a very broad range of particle energies \citep[see e.g.][]{drury2017,gabici2019}.

The SNR W28 is a middle-aged remnant (estimated age equal to few times $10^4$ years) located at a distance of about 2 kpc \citep{velazquez2002}.
It is classified as a mixed-morphology SNR with center-filled thermal X-ray emission and shell-like radio morphology \citep{rho2002,dubner2000}.
Also, observations in CO(1-0) have revealed molecular gas within the field of W28 \citep{dame2001}, concentrated in a number of massive Molecular Clouds (MCs) \citep{aharonian2008}. Most importantly the detection of 1750 OH maser from the MC located on the north-eastern side of the SNR suggests that this cloud is interacting with the blast wave of the remnant \citep{claussen1997}.
The W28 SNR/MC system has been observed at all wavelengths, including radio \citep[see][and references therein]{dubner2000}, millimeter \citep{vaupre2014}, X-rays \citep[see][and references therein]{zhou2014}, high energy and very high energy gamma rays \citep{aharonian2008,abdo2010}.
This makes it an ideal target for studies of CR acceleration and escape from SNRs \citep[e.g.][and references therein]{nava2013}.

The MCs in the vicinity of the SNR W28 are prominent gamma-ray sources \citep{aharonian2008,abdo2010}. The origin of this emission is due to interactions of GeV and TeV CR protons that were accelerated in the past at the SNR shock and that now fill a vast region surrounding the remnant \citep[e.g.][]{gabici2010}.  
Remarkably, measurements performed in the millimeter domain revealed an enhanced ionization rate from the north-eastern MC \citep{vaupre2014}. 
The ionization of the MC could be due either to the interactions of CRs (either protons or electrons) in the molecular gas, or to the presence of X-rays coming from the SNR shock-heated gas. X-rays have been proposed as a possible source of ionization in the vicinity of a number of SNRs \citep{schuppan2014}.

As we will see in the following, X-rays are in this case not a viable explanation for the enhanced ionization, and therefore CRs are left as the only possible ionizing agents present inside of the cloud.
This fact opens the possibility to combine high and low energy observations of the SNR/MC system (gamma rays and millimeter waves, respectively), and constrain the spectrum of CRs present in the region over an interval of particle energies of  unprecedented breadth: from the MeV to the TeV domain. We will show that data are best explained if an enhanced flux of CR protons is present, and if such protons are characterized by energies spanning from $\lesssim 100$ MeV up to tens of TeV.

The paper is organized as follows: in Section 2 we summarize the multi-wavelength observations of the W28 SNR/MC system, in Section 3 we compute the photoionization rate induced in the MC by X-ray photons. 
The role of CRs in ionizing the gas is investigated in Section 4, where constraints on the CR proton and electron specxtra are also obtained.
We discuss our results and conclude in Section 5.

\section{Multi-wavelength observations of the W28 region}

\begin{figure}[h]
\includegraphics[width=3.5in]{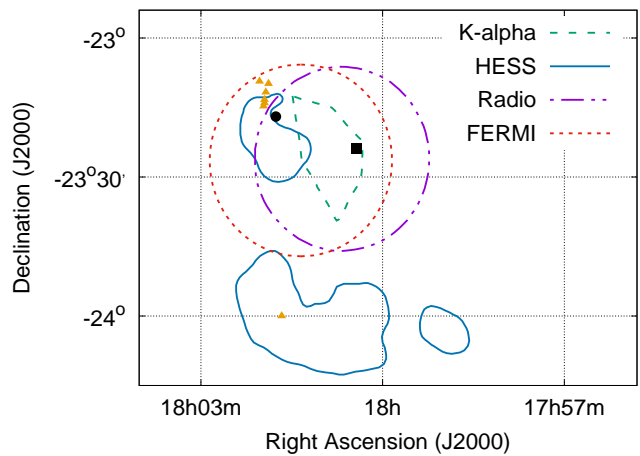}\label{fg_position}
\caption{Contour map for the W28 region. The approximate radio boundary of the SNR shell is shown as a purple dot-dot-dashed circle. The solid blue contours represent the 4$\sigma$ significance excess TeV emission observed by H.E.S.S. \citep{aharonian2008}. The short-dashed red circle is the best-fit disk size for the \textit{Fermi}-LAT GeV source associated to the north-eastern MC \citep{cui}.
The enhanced region of Fe $\text{\rom{1}}$ K$\alpha$ line emission is the area enclosed by the dashed green line \citep{nobukawa2018}.
The CR ionization rate has been measured from IRAM observations pointed in the directions indicated by the yellow triangles \citep{vaupre2014}. The filled black circle and square indicate the centroids of the X-ray emission for the North-East and Central X-ray sources, respectively \citep{rho2002}.}\label{fig:contour}
\end{figure}

In this Section we review the status of the multi-wavelength observations of the SNR W28 and its surroundings. The purple (dot-dot-dashed) circle in Fig.~\ref{fig:contour} indicates the approximate contours of the SNR shell, as traced by its radio emission \citep{dubner2000,bogan2006}. Observations in the CO molecular line revealed the presence of a number of dense ($\approx 10^3$ cm$^{-3}$) and massive ($\approx 10^5 M_{\odot}$) MCs in the region \citep{matsunaga2001,aharonian2008}. Remarkably, the H.E.S.S. collaboration reported on the detection of very-high-energy gamma-ray emission from the vicinity of W28, which correlates spatially very well with the position of the MCs \citep{aharonian2008}. The blue contours in Fig.~\ref{fig:contour} show the 4$\sigma$ significance excess in TeV gamma rays. The spatial correlation points towards an hadronic origin of the gamma-ray emission, which thus results from the interactions of CR nuclei with the dense gas that forms the MCs. Gamma-ray data are best explained by assuming that CR protons were accelerated in the past at the SNR, when the shock speed was larger than the present one. Such particles then escaped the system, and now fill a large volume which encompasses all the gamma-ray bright MCs \citep{fujita2009,gabici2010,li2010,ohira2011,nava2013}. 

The detection of OH maser emission from the north-eastern MC indicates that the SNR shock is currently interacting with that cloud \citep{claussen1997,hewitt2008}. The other TeV-bright MCs are located in the south, outside of the SNR radio boundary and therefore have not been reached by the shock yet. In the following, we will focus mainly on the interaction region, and for this reason we also show, as a dashed red circle, the position and extension of the \textit{Fermi}-LAT source associated to the north-eastern cloud \citep{abdo2010,cui}.

The presence of the gamma-ray emission from the MCs reveals an overdensity of CRs with respect to the Galactic background, both in the GeV and TeV energy domain. In addition to that, observations of millimetre lines performed with the IRAM 30 metres telescope (orange triangles in the Figure) showed that an excess in the gas ionization rate is also present at the position of the SNR/MC interaction, but not at the position of the southern MC complex \citep{vaupre2014,gabici2015}. Such enhanced ionization rate could be interpreted as an excess of CRs (either protons or electrons) of low energy ($\approx$ MeV energy domain). However, the SNR is a powerful thermal X-ray source \citep{rho2002,zhou2014}, and the X-ray photons might also penetrate the cloud and be responsible for the enhanced ionization rate, as it was proposed for other SNR/MC systems \citep{schuppan2014}. 
The spatial morphology of the X-ray emission is quite extended, and can be roughly described as the sum of two extended sources, whose centroids are shown in Fig.~\ref{fig:contour} as a filled black square (central source, C) and circle (north-eastern source, NE).
Determining whether the enhanced ionization rate is due to CR protons, electrons, or X-ray photons is one of the goals of this paper (see Sec.~\ref{sec:photoionisation}). 

Finally, additional constraints on the origin of the enhanced ionization rate can be obtained from hard X-ray observations of W28 performed by {\it Suzaku} \citep{nobukawa2018}. These observations revealed the presence of the Fe I K$\alpha$ line in the X-ray spectrum. This line is produced by interactions between low energy (MeV domain) CRs and cold gas, and it is therefore tempting to propose a common origin for the line emission and the excess in the ionization rate measured in the north eastern MC. Puzzingly, Fe I K$\alpha$ line emission has been detected from a region (green dashed contour in Fig.~\ref{fg_position}) close but not coincident with the position of the gamma-ray bright north eastern MC.

\section{Photoionization}
\label{sec:photoionisation}

Based on XMM-Newton observations, \citet{zhou2014} claimed that the X-ray emission from the SNR W28 is predominantly thermal, with a possible sub-dominant non-thermal contribution from source NE, and an indication for the presence of a multi-temperature gas for source C (see also \citealt{rho2002}).

In order to estimate the level of photoionisation induced by the SNR X-ray emission in the north-eastern MC, we make use of the spectral fits to XMM data obtained by \citet{zhou2014}. For simplicity, we consider single-temperature non-equilibrium ionisation models ({\it vnei} model in XSPEC\footnote{\url{https://heasarc.gsfc.nasa.gov/xanadu/xspec/}}) to describe the X-ray emission of both the NE and C source, and we provide the best fit parameters in Table~\ref{tab:X}. Even though adding a power law component to the spectrum of the NE sources and a second thermal component to the C source somewhat improves the spectral fits, this would have no major effect in our estimate of the photoionisation rate (this can be easily tested {\it a posteriori}).


\begin{table}
\centering
\caption{Fit parameters of \textit{vnei} model for the X-ray sources adopted from \cite{zhou2014}}

	\label{tab-xray}
	\begin{tabular}{lcr} 
		\hline
		Objects$\vphantom{^{\frac{•}{•}}}$ & NE & C  \\
		\hline
		$kT_c$ (keV) $\vphantom{^{\frac{•}{•}}}$ & 0.33  & 0.60 \\
		$\tau_c$ ($10^{11}$ cm$^{-3}$s) & 6.00 & 2.35\\
		$^1$Abun  & 0.26 & 0.12\\ 
		$\text{[Si/H]}$ & 0.40 & 0.16 \\
		$\text{[S/H]}$ & 0.88 & 0.39 \\
		$\text{[Fe/H]}$ & 0.22 & 0.10 \\
		$^2 \phi_s$ ($10^{-11}$ erg cm$^{-2}$ s$^{-1}$) & 6.61 & 10.43 \\
		\hline
		 \multicolumn{3}{l}{\scriptsize  $1.$ Abundances of C, N, O, Ne, Mg, Ar, Ca, and Ni.}\\
   \multicolumn{3}{l}{\scriptsize  $2.$ Unabsorbed fluxes in the 0.3–5.0 keV band.}\\
	\end{tabular}
	\label{tab:X}
\end{table}



\begin{figure}[h]
\includegraphics[width=3.5in]{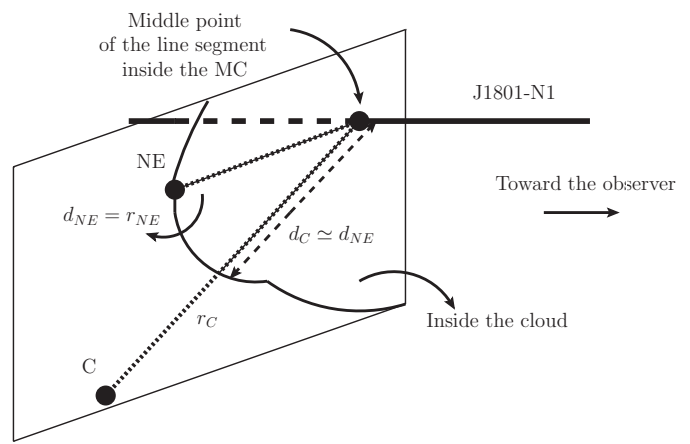}\label{fg_w28}
\caption{Adopted geometry for the calculation of photoionisation. The centroids of the X-ray emission are indicated by the points labeled NE and C, the line of sight labeled J1801-N1 is the IRAM pointing closest to the X-ray sources.}\label{fig:cartoon}
\end{figure}
\begin{figure*}[ht]
\label{fig:zeta}
\centering
\includegraphics[width=3.5in]{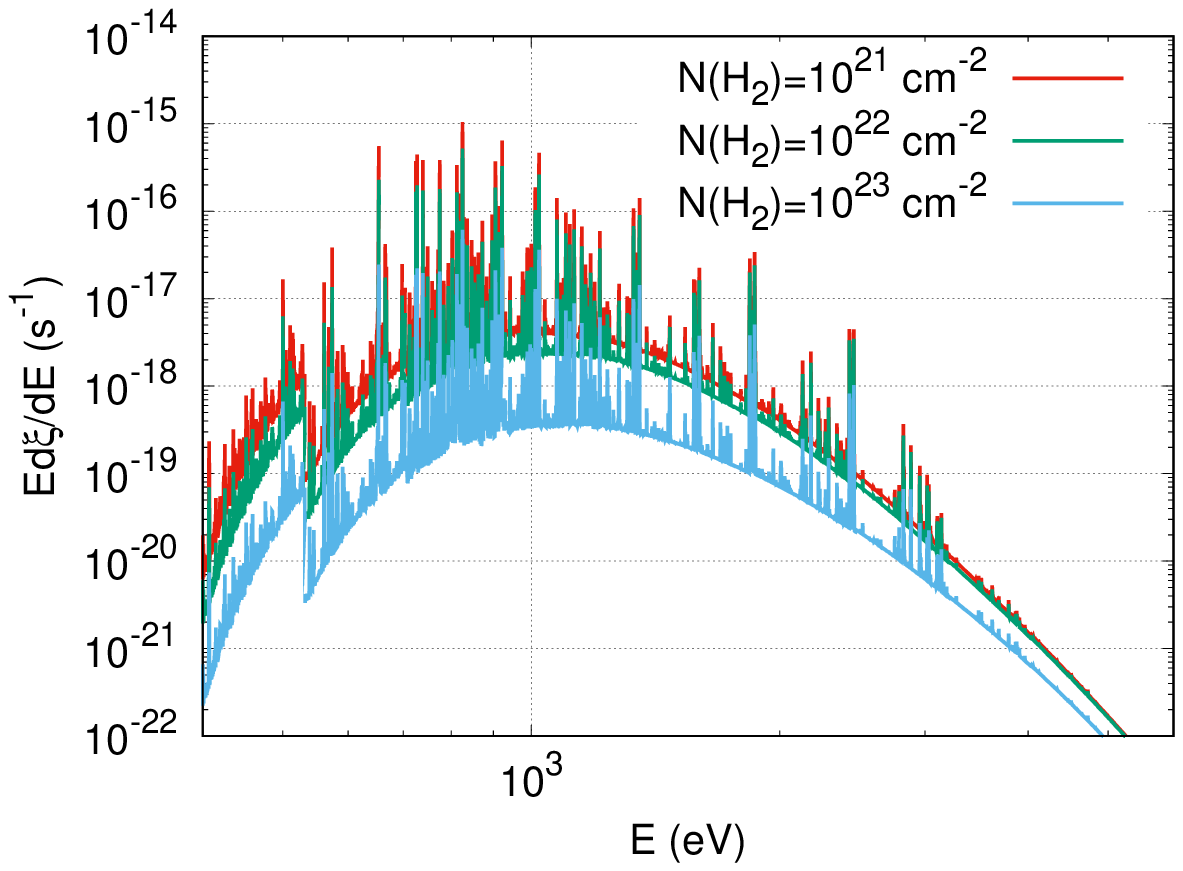}
\includegraphics[width=3.5in]{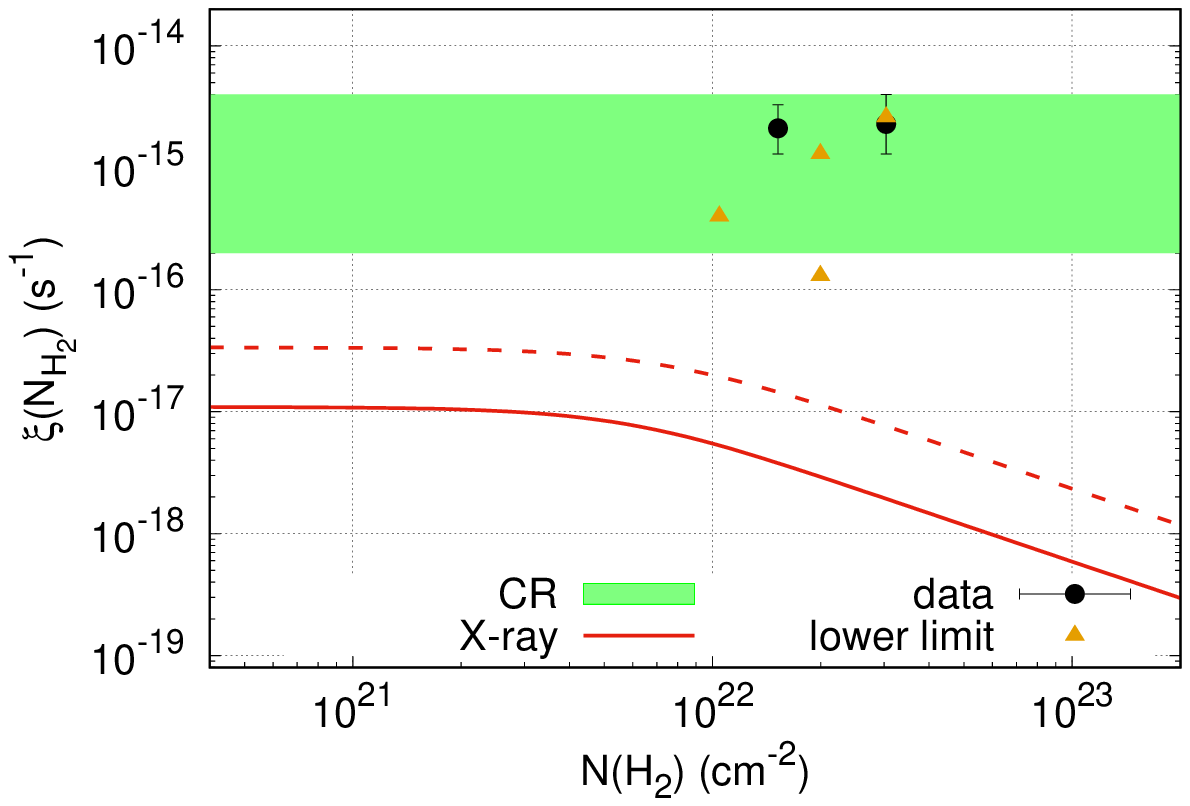}
\caption{\textit{Left}: Average differential ionization rate for different assumed gas column densities along the line of sight N1 and solar abundance. \textit{Right}: Predictions for the photo-ionization rate are shown by the solid and dashed red lines (metallicity of 1 and 0.3 times the solar one, respectively). Observational data taken from \cite{vaupre2014} are presented as filled circles (measurements) and filled triangles (lower limits). The shaded region indicates a reference range of values of the ionization rate which has been used to constrain the CR spectrum (see text). }\label{f4}
\end{figure*}

The intensity of the X-ray radiation inside the north-eastern MC has been computed by assuming that the X-ray sources NE and C are point like and located at the position of the two centroids of the X-ray emission (filled black circle and square in Fig.~\ref{fig:contour}). The contribution to the density of X-ray photons at a given location $\mathbf{r}$ away from the source can be computed as:
\begin{eqnarray}
n_{ph}(E,\mathbf{r})=\frac{F(E)D_s^2}{cr^2}\exp\left[-\frac{n_{\text{H}_{2}}  \sigma_{abs}(E) }{f_{\text{H}_{2}}}d(r) \right]\label{eq1}
\end{eqnarray}
where $F(E)$ is the X-ray source's unabsorbed differential photon flux, $D_s \sim$ 2 kpc is the distance from the source to Earth, $c$ is the speed of light, $n_{\text{H}_2}$ is the density of H$_2$ molecules in the absorbing medium (which in this case is the MC), $f_{\text{H}_2}=n_{\text{H}_2}/(2n_{\text{H}_2}+n_{\text{H}})$ is the fractional density of H$_2$ molecules relative to the total number of H atom, $\sigma_{abs}(E)$ is the photoelectric absorption cross-section per H atom, and $d(r)$ is the distance travelled inside of the cloud by the X-ray photons that reached a distance $r$ from the source (see Fig.~\ref{fig:cartoon}).
For MCs, it is appropriate to set $f_{\text{H}_2} \sim 0.5$ \citep[see e.g.][]{vaupre2014}, and we further assume solar abundances to describe the gas in the MC. This latter assumption allows us to use the absorption cross-section $\sigma_{abs}(E)$ taken from \citet{morrison}. In fact, the exact value adopted for the element abundances in the MC might have some impact: changing the abundances from 1 to $\sim$ 0.3 times the solar one (i.e. close to the abundance of the north-eastern MC as reported in Table~\ref{tab:X}) would increase by a factor of $\sim 3$ the estimate of the photoionisation rate (see Fig.~\ref{f4}).

With Eq.~\ref{eq1} at hand, the photoionisation rate $\xi_{ph}$ induced by X-ray photons inside the cloud can be obtained following the approach presented in \citet{maloney1996}. Due to the relatively low temperatures of the emitting plasmas ($\approx 1$ keV), the contribution from Compton scattering to the photoionisation rate can be safely neglected. Moreover, most of the ionization in the MC will be induced by secondary electrons generated as a result of the X-ray photoionisation, so that we can write:
\begin{eqnarray}
\xi_{ph}(\mathbf{r})=2f_{\textrm{H}_2}\sum_{s}\int^{E_{max}}_{I(\text{H}_2)}\sigma_{abs}(E)cn^{s}_{ph}(E,\mathbf{r})M_{sec}(E)\df E\label{eq2}
\end{eqnarray}
where the sum indicates that both X-ray sources ($s =$ NE and C) are considered.
Here, $I(\text{H}_2)\approx 15.6$ eV is the ionization potential of H$_{2}$, and $M_{sec}(E)=[E-I(\text{H}_2)]/W$ is the mean multiplicity for ionisation by a secondary electron in a $H_2$ gas, with $W \sim 40$ eV \citep[see][]{dalgarno1999,dogiel2013}.

Before proceeding, a discussion of the geometry of the problem is in order (see Fig.~\ref{fig:cartoon}). We model the X-ray emission from the SNR as two point sources located at the position of the centroids of the emission \citep{rho2002}. The ionisation rate in the north-eastern MC has been measured along several line of sights (yellow triangles in Fig.~\ref{fig:contour}, \citealt{vaupre2014}). To maximise the effect of photoionisation, we have chosen to study the line of sight which is the closest to the centroids of the X-ray emission, which has been labeled by \cite{vaupre2014} as J1801-N1. This line of sight is also the one characterised by the smaller measured value of the density of $H_2$, which is equal to $n_{H_2} \sim 600$ cm$^{-3}$ \citep{vaupre2014}. In the following we assume that this density characterises the entire cloud. This is of course not true, given that larger values of this quantities has been estimated by \citet{vaupre2014} for all the line of sights other than J1801-N1, but such an assumption will provide us with the most optimistic (larger) estimate of the photoionisation rate in the cloud. We further assume then the two X-ray sources to lay on a plane orthogonal to the line of sight J1801-N1, and the north-eastern MC to be spatially symmetric with respect to that plane.
The coordinates of the two X-ray sources and of the line of sight J1801-N1 are listed in Table~\ref{tab-position}, together with their relative distances. 

\begin{table}
\centering
\caption{Coordinates and relative distances of the two X-ray sources and of the line of sight J1801-N1.}

	\label{tab-position}
	\begin{tabular}{lcccr} 
		\hline
		\multirow{2}{*}{Objects}& $\alpha$ $\vphantom{^{\frac{•}{•}}}$ & $\delta$ & $d_{NE}$ & $d_C$ \\
        & (hms) & ($\degree$ ' '') & (pc) & (pc)$\vphantom{_{\frac{•}{•}}}$ \\
		\hline
		NE source $\vphantom{^{\frac{•}{•}}}$ & 18 01 45.7  & -23 16 58.3 & $-$ & $-$  \\
		C source & 18 00 25.5 & -23 23 47.2 & $-$ & $-$ \\
		J1801-N1 & 18 01 58.0 & -23 14 44.0 & 2.10 & 13.43 \\ 
		\hline
	\end{tabular}
\end{table}

Adopting the geometry presented in Fig.~\ref{fig:cartoon}, we can now estimate the X-ray photo-ionization rate as a function of the gas column density along the line of sight $N(H_2)$. Figure \ref{f4} shows the differential photo-ionization rate averaged along the line of sight for different values of the column densities along the line of sight: $N({\text{H}_2})=10^{21}$, $10^{22}$, $10^{23}$ cm$^{-2}$, respectively.
For typical values in the range $N(H_2) \approx 10^{21}$ - $10^{23}$ cm$^{-2}$, Fig.~\ref{f4} (left panel) shows that the contribution to ionization comes mainly from X-ray photons with energy in the range from $\lesssim$ 1 keV to few keV.

The right panel of Fig.~\ref{f4} shows the expected photo-ionization rates as a function of the gas column density for a cloud of solar (solid red curve) and 0.3 times solar metallicity (dashed red curve). These predictions can be compared with the measurements of the ionization rate in the north eastern cloud (data points and lower limits in the Figure). 
One can clearly see that the contribution from X-rays to the observed ionization rate is negligible even in the most optimistic scenario considered here. Therefore, the ionization rate must be due to CRs, either protons (or nuclei) or electrons. We will consider this scenario in the next Section.

\section{Cosmic-ray induced ionization}

The gamma-ray emission detected from the MCs in the vicinity of the SNR W28 is interpreted as the result of hadronic interactions between CR nuclei (mostly protons) accelerated in the past at the SNR shock. These particles escaped the remnant and now fill a vast region that encompasses the clouds. The gamma-ray emission results from the decay of neutral pions produced in inelastic proton-proton interactions. The energy threshold to produce pions at rest is $T_p^{th} \sim 280$ MeV. Therefore, gamma-ray observations can be used, generally, to determine the shape of the spectrum of CR protons of kinetic energy exceeding $T_p^{th}$ contained in the MCs.

It is definitely less straightforward to infer the energy spectrum of the CR electrons contained within the cloud. The SNR W28 is a bright synchrotron radio source \citep{dubner2000}, and this indicates the presence of relativistic electrons and magnetic field in the region. Even though the morphology of the radio emission does not correlate with that of molecular clouds, the brightest region in radio roughly coincides with the position of the north eastern cloud.

In the following, we will investigate the possibility that either CR protons or electrons are the responsible for the enhanced ionization rate measured from the north-eastern cloud.


\subsection{Cosmic ray protons and nuclei}
\label{sec:protons}

Let us assume that the spectrum of CR protons in the cloud can be described as a power law in momentum $\propto p^{-(\delta_p+2)}$ as expected if protons are accelerated at the SNR shock via first order Fermi mechanism.
In terms of the particle kinetic energy $T_p$ this writes:
\begin{equation}
\label{eq:CRp}
 n_p(T_p) = A_p \left( T_p+m_p c^2\right) \left[ T_p^2 + 2 T_p m_p c^2 \right]^{-\frac{\delta_p+1}{2}}  
\end{equation}
where all energies are in GeV and $A_p$ is a normalisation factor. Both $A_p$ and the spectral index $\delta_p$ can be obtained by fitting the gamma-ray data.

\begin{figure*}[ht]
\centering
\includegraphics[width=3.5in]{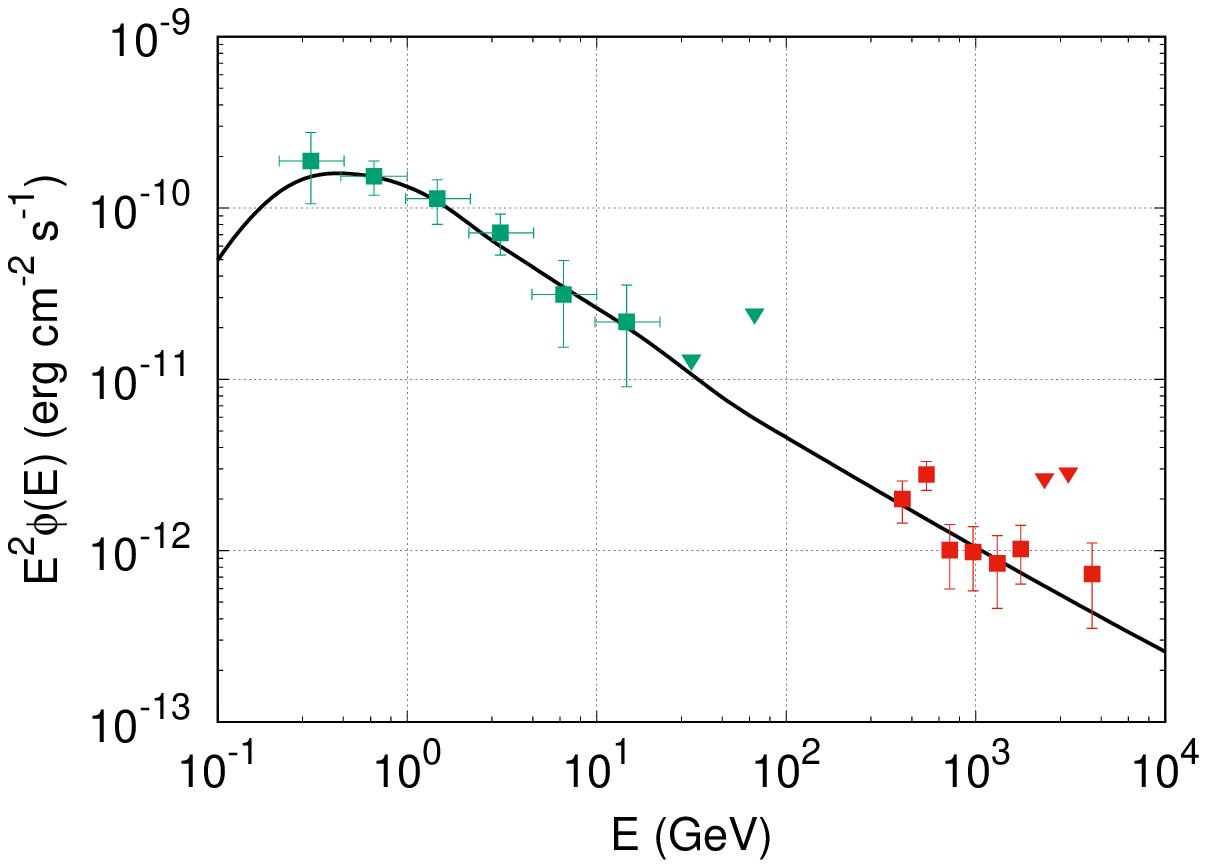}
\includegraphics[width=3.5in]{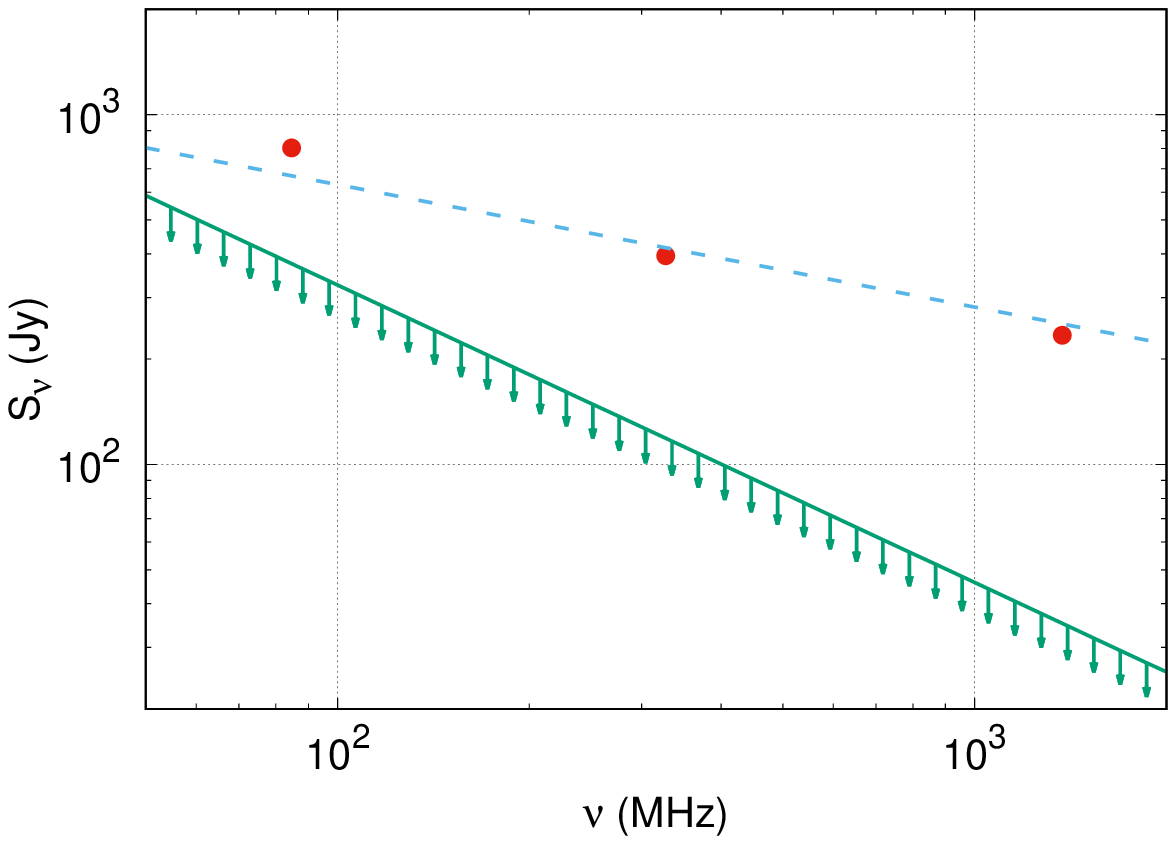}
\caption{\textit{Left}: Gamma-ray differential energy spectrum of the northeastern MC (J1801-233). \textit{Fermi}-LAT and HESS data have been fitted with an hadronic model \cite[see][for references]{aharonian2008,abdo2010}. \textit{Right}: Synchrotron emission spectrum  observed from the entire SNR \citep{kovalenko,dubner2000}. A fit to data is shown as a dashed line. The maximum contribution from the north-eastern cloud is also shown (see text for details). }\label{f5}
\label{fig:spectra}
\end{figure*}

Assuming that the observed gamma-rays are produced by proton-proton interactions
, the expected gamma-ray flux measured at Earth would be:
\begin{eqnarray}
\phi(E_\gamma)&=&\frac{M_{cl}}{4\pi D_s^2 m_{avg}}\int^{T^{\text{max}}_{p}}_{T_p^{min}}  4\pi J_p(T_p)\n\\
&&\qquad \qquad \qquad \times \varepsilon(T_p)\frac{\df \sigma_{pp}(T_p,E_\gamma)}{\df E_\gamma}\df T_p
\end{eqnarray}
where $J_p(T_p) = (v/4 \pi) n_p(T_p)$ is the CR proton intensity as a function of the particle kinetic energy $T_p$ ($v$ is the particle velocity). Moreover, $M_{cl}$ is the mass of the whole cloud ($\sim 5 \times 10^4 M_{\odot}$, see \citealt{aharonian2008}), $m_{avg}$ is the average atomic mass of the gas from the MC (with solar abundance $m_{avg}\simeq 1.4m_p$), $T^{\text{max}}_{p}$ is the maximum kinetic energy of the accelerated particle (its exact value is irrelevant, as long as $\gg$ 100 TeV), and $T_p^{min}$ is the threshold energy for $\pi^0$ production in proton-proton interactions. Also, $\df \sigma_{pp}(T_p,E_\gamma)/\df E_\gamma$ is the differential cross-section for gamma-ray production  and $\varepsilon(T_p)$ is the nuclear enhancement factor to take into account gamma-ray production from nucleus-nucleus interaction \citep[both taken from][]{kafexhiu2014}.  A fit to the gamma-ray data is shown in Fig.~\ref{fig:spectra} (left panel), where data points are from \citet{aharonian2008} and \citet{abdo2010}. The values obtained for $A_p$ and $\delta_p$ are reported in the first row of Table \ref{tab-fit}. Of course, the CRs responsible for the gamma ray emission are characterised by particle energies in the GeV and TeV domain, therefore their contribution to the ionization rate is negligible \citep[e.g.][]{padovani2009,phan2018}.


In order to estimate the possible contribution from CR protons to the measured ionization rate in the cloud, we extrapolate the power low spectrum obtained after fitting the gamma-ray data down to the MeV energy domain. In other words, we assume a spectrum as in Eq.~\ref{eq:CRp} down to an arbitrary particle kinetic energy $T_c$. 
We compute then the CR ionization rate following \citet{krause} \citep[see also][]{phan2018,recchia2019}, and we constrain the value of $T_c$ so that the ionization rate falls in the range indicated by the green shaded region in Fig.~\ref{fig:zeta}. The range of values $T_c^{min} < T_c < T_c^{max}$ obtained in this way are reported in Table \ref{tab-fit}.

There are several reasons to envisage a change in the CR proton spectrum at certain particle energies $T_c$. Let's first consider the scenario where the region of enhanced ionization is upstream of the shock and the ionizing CRs have already escaped the remnant. In this case, the range of possible numerical values of $T_c$ can be estimated as follows.
\begin{enumerate}
    \item{CR protons have been produced at the SNR shock a time $\tau_{inj}$ ago, and since then they suffer energy losses (mainly ionization losses) in the dense gas, over a characteristic time $\tau_{ion}(T_p)$, which is proportional to the gas density and an increasing function of particle energy (see Fig. 2 in \citealt{phan2018}).
    In fact, energy losses can be effective only for particles characterised by an energy smaller than $T_c$, defined as $\tau_{ion}(T_c) = \tau_{inj}$, because particles of higher energy simply do not have time to cool. The maximum possible value for $\tau_{inj}$ is of course the age of the SNR $\tau_{age} \approx 4 \times 10^4$ yr \citep{gabici2010}, which provides an upper limit for $T_c$. For a typical gas density of $n_{{\rm H}_2} \sim 10^3$ cm$^{-3}$ this gives 
    $
        T_c \lesssim 4 \times 10^2 {\rm MeV}
    $, which is quite close to the value of $T_c^{max}$ in Table \ref{tab-fit}.
    }
    \item{CR protons have to penetrate deep into the cloud in order to ionize the gas there. If we call $\tau_p$ the time it takes them to reach the centre of the cloud moving a distance $L$ away from the position of the shock, we can estimate $T_c$ by imposing $\tau_{ion} = \tau_p$. The shortest possible penetration time $\tau_p$ is obtained after assuming that CR protons move along straight lines at a velocity $v \sim \sqrt{2 T_c/m_p}$. If we assume that CR protons have to cross a gas column density of $\sim 3 \times 10^{22} (n_{{\rm H}_2}/10^3 ~ {\rm cm}^{-3}) (L/10~{\rm pc})$ cm$^{-2}$ we get $T_c \gtrsim 7$ MeV, which is a factor of few smaller than $T_c^{min}$ in Table \ref{tab-fit}.}
\end{enumerate}

It should be noticed that, in this scenario, the range of possible values for $T_c$ obtained by means of the phenomenological consideration made above overlaps very well with the range of values obtained observationally (Table \ref{tab-fit}), i.e., by fitting simultaneously the millimeter and gamma-ray data.

Another possible scenario is that the SNR shock has overrun the region of enhanced ionization, engulfing it with low energy CRs which are still inside the shell, in the downstream region. This is similar to the case of the region of enhanced ionization W51C-E which has been shown to be in the downstream region of the SNR W51C (see \citealt{dumas2014} for more detailed discussion). If this is true then we may expect a hardening in the low energy part of the spectrum, below $T_c$, due to the difference in spectral features of the escaped CRs at high energy and the still confined CRs at low energy. However, our understanding of the escape of CRs from SNR shocks is still quite poor, making an accurate estimate of the numerical value of $T_c$ problematic.

Independently on the scenario, the important point that needs to be stressed is that gamma-ray observations allow us to constrain the spectrum of CR protons for particle energies above $\approx 1$ GeV. Therefore, the explanation of the enhanced ionization rate requires to extrapolate the proton spectrum only by a factor of $\approx 3-30$ down to lower particle energies. 

It follows that the presence of an excess of CR protons characterized by a quite steep power law spectrum ($\delta_p \approx 2.8$) extending from the MeV to multi-TeV domain can explain very naturally both the bright gamma-ray emission from the north eastern cloud and the observed enhancement in the ionization rate.

\begin{table}
\centering
\caption{Fit parameters for the CR proton spectrum and upper limit for the CR electron spectrum.}

	\label{tab-fit}
	\begin{tabular}{lccc} 
		\hline
		Species$\vphantom{^{\frac{•}{•}}}$ & $A_{p,e}$ & $\delta_{p,e}$ & $T_c^{\text{min}}-T_c^{\text{max}}$ \\
		$\vphantom{^{\frac{•}{•}}}$ & (eV$^{-1}$ cm$^{-3}$) &  & (MeV)\\
		\hline
		Proton $\vphantom{^{\frac{•}{•}}}$ & $3.15\times 10^{-17}$  & $2.76$ & $26-320$ \\
		Electron &  $\ll 6.4\times 10^{-19}$ & $2.7$ & $\ll 20 - 130$ \\
		\hline
	\end{tabular}
\end{table}

\subsection{Cosmic ray electrons}

As stated above, the decay of neutral pions produced in inelastic interactions between CR protons (and nuclei) and the dense gas in the cloud provides the most natural explanation for the gamma-ray emission observed in the GeV and TeV energy domain \citep[see e.g.][]{aharonian2008,nava2013}. Leptonic models where the emission is due to non-thermal Bremsstrahlung have been shown to be problematic, as they require the SNR to accelerate the same amount of nuclei and electrons, and provide a good fit to gamma-ray data only if unrealistic values of the local magnetic field strength and gas density are assumed \citep{abdo2010}.

This implies that the non-thermal Bremsstrahlung emission from CR electrons must provide a sub-dominant contribution to the observed gamma-ray emission.
If $n_e(E_e) = A_e (E_e/{\rm GeV})^{-\delta_e}$ is the density of relativistic electrons of energy $E_e$ inside the cloud, the expected Bremsstrahlung emission can be roughly estimated as:
\begin{equation}
E_{\gamma}^2 L_B(E_{\gamma} \sim E_e) \sim \frac{E_e^2 n_e(E_e) V_{cl}}{\tau_B}    
\end{equation}
where
\begin{equation}
\tau_B \sim 4 \times 10^7 \left( \frac{n_H}{\rm cm^{-3}} \right)^{-1} ~ \rm yr    
\end{equation}
is the energy loss time due to Bremsstrahlung in a gas characterised by an atomic Hydrogen density $n_H$ \citep{aharonian2004} and $V_{cl}$ is the volume of the cloud.

By imposing that the Bremsstrahlung flux $E_{\gamma}^2 L_B/(4 \pi D_s^2)$ should be significantly smaller than the observed one $E_{\gamma}^2 \phi(E_{\gamma})$ we get:
\begin{equation}
E_e^2 n_e(E_e) \ll 4 \pi D_s^2 \left( \frac{\tau_B}{n_H}  \right) E_{\gamma}^2 \phi(E_{\gamma}) m_p M_{cl}^{-1}
\end{equation}
where we introduced the source distance $D_s$ and the cloud mass $M_{cl} = m_p n_H V_{cl}$.
At a photon energy equal to $E_{\gamma} = 1$ GeV the observed gamma-ray flux is roughly $E_{\gamma}^2 \phi(E_{\gamma}) \approx 10^{-10}$ erg/cm$^2$/s, and above such energy the spectrum can be described by a power law $E_{\gamma}^{-\alpha}$ with $\alpha \sim 2.7$ \citep{abdo2010,aharonian2008}.
This gives:
\begin{equation}
\label{eq:upper}
E_e^2 n_e(E_e) \ll 10^{-12} \left( \frac{E_e}{\rm GeV} \right)^{-0.7} ~ \rm erg/cm^3
\end{equation}
which corresponds to $A_e \ll 6.4 \times 10^{-19}$ eV$^{-1}$ cm$^{-3}$.

We note that the ratio between the intensity of CR protons and electrons inside the cloud is $A_p/A_e \gg 50$ for particle energies of the order of 1 GeV, and varies very little for larger particle energies (because $\delta_p \sim \delta_e$).

The constrained obtained above on the electron spectrum in the cloud can be used to estimate the contribution given by such electrons to the observed radio emission from the SNR.
Radio observations of the SNR W28 have been performed in a frequency range spanning from $\lesssim 100$ MHz to several GHz. The radio spectrum is shown in the right panel of Fig.~\ref{fig:spectra}: the flux at 1.4 GHz is $S_{1.4} \sim 246$ Jy, and the radio spectrum can be described by a power law $S_{\nu} \propto \nu^{-\alpha}$ with $\alpha \sim 0.35$ \citep{dubner2000}. Even though the radio emission is observed from a region which is spatially more extended than the northern cloud, the radio brightness roughly peaks at that position. Therefore, it is natural to ask whether a contribution to the radio emission might come from relativistic electrons located inside the cloud.

Electrons of energy $E$ emit synchrotron photons of frequency \citep{aharonian2004}:
\begin{equation}
\nu_s \sim \nu_c/3 = \frac{\Omega_L}{2} \left( \frac{E_e}{m_e c^2} \right)^2 \sim 0.2 \left( \frac{B}{40~\mu {\rm G}} \right) \left( \frac{E_e}{\rm GeV} \right)^2 \rm GHz
\end{equation}
where $\nu_c$ is the critical synchrotron frequency, $\Omega_L$ the non-relativistic Larmor frequency, and $m_e c^2$ the electron rest mass energy.
This implies that the observed radio emission is produced by electrons of energies in the range $0.1 \lesssim (E_e/{\rm GeV}) (B/40~\mu {\rm G})^{1/2} \lesssim 10$. 
Electrons of such energies emit Bremsstrahlung in the Fermi energy domain.

The value of the magnetic field in the cloud $B$ has been normalized to what is expected from the observational relationship $B \approx 10 (n/100~{\rm cm}^{-3})^{0.5} \mu$G between the cloud magnetic field and the density of the gas \citep{crutcher}. For the north-eastern cloud, \citet{aharonian2008} estimated a gas density equal to $n \sim 1.4 \times 10^3$ cm$^{-3}$, which would give $B \sim  37 \mu$G.

The upper limit on the CR electron spectrum obtained above (Eq.~\ref{eq:upper}) can now be used to estimate the maximum contribution to the observed synchrotron emission coming from the north-eastern cloud. This is shown in the right panel of Fig.~\ref{fig:spectra}. The contribution is subdominant at all frequencies. This implies that the radio emission is largely produced by electrons located outside of the cloud. 

It is not at all straightforward, then, to estimate the possible contribution of CR electrons to the ionization rate in the cloud. Even though we do not know the spectral slope $\delta_e$ and normalization $A_e$ of the CR electrons inside the MC, we have obtained an upper limit for the latter value from gamma-ray observations (see Table~\ref{tab-fit}) which
applies to electrons of particle energy above $E_e^{min} \approx E^{min}_\gamma \approx 0.2$ GeV. 
Electrons of energy larger than $E_e^{min}$ and characterized by an intensity well below the upper limit reported in Eq.~\ref{eq:upper} would produce an ionization rate of the order of $\ll 1.4 \times 10^{-16}$ s$^{-1}$ (following also the procedure in \citealt{krause}). 
Therefore, as one can easily see from Fig.~\ref{fig:zeta}, electrons of energy exceeding $E_e^{min}$ cannot explain the observed ionization rate in the cloud.

An extrapolation of the \textbf{upper limit} electron spectrum to much lower particle energies might possibly result in a value of the ionization rate comparable with observations. Using again the upper limit for $A_e$, together with a spectral index of $\delta_e\simeq 2.7$, which is quite steep and therefore maximises the impact of electrons, we obtain that values of $T_c$ well below 20--130 MeV are needed in order to match the observed ionization rate (see Table~\ref{tab-fit}). However, the validity of such an extrapolation might be questionable since the spectral index for electrons is not as well constrained as that of protons.

Even though we cannot reach a firm conclusion, explaining the excess in the ionization rate with CR protons is very natural (Sec.~\ref{sec:protons}) and therefore such hadronic scenario remains, in our view, to be preferred.

\subsection{Constraints by the 6.4 KeV line emission}

Interestingly, the SNR W28 is also observed by \textit{Suzaku} in the Fe \rom{1} K$\alpha$ line emission and it is found quite recently by \cite{nobukawa2018} that there is an excess of of this emission from the central region of the remnant with intensity $I^{en}_{6.4\text{keV}}=(3.14\pm 0.43)\times10^{-8}$ cm$^{-2}$s$^{-1}$arcmin$^{-2}$ compared to the background of the Galactic ridge X-ray emission $I^{bg}_{6.4 \text{keV}}=(2.33\pm 0.29)\times10^{-8}$ cm$^{-2}$s$^{-1}$arcmin$^{-2}$. The authors concluded that this excess emission is produced due to the interaction of of MeV CR prorons from the SNR with the ambient cold gas. 
It seems, however, quite puzzling since this explanation would require a relatively high column density of the gas, comparable to that of the molecular cloud in the northeastern part of the remnant ($N_{\text{H}_2}\simeq 10^{22}$ cm$^{-2}$). However, the region of enhanced emission as defined in \cite{nobukawa2018} and the TeV gamma-ray contour by HESS (respectively the area enclosed with the dashed blue line and the green contour in Fig.~\ref{fg_position}) do not overlap. 
We conclude, then, that the Fe \rom{1} K$\alpha$ line emission must have a different origin.

Nevertheless, the Fe \rom{1} K$\alpha$ line emission is quite an informative channel for the study of low energy CRs and could also be used to put some constraints on the CR spectra. In particular, the total flux coming from the excess measured by \citet{nobukawa2018} can be computed as $F_{ex} = \Delta \vartheta (I^{en}_{6.4\text{keV}} - I^{bg}_{6.4 \text{keV}})$ where $\Delta \vartheta \approx 15^2$ arcmin$^2$ is the extension of the excess region. Such a flux, if produced by CR protons in the north-eastern cloud, would have been detected.

The non-detection of the iron line from the north-eastern cloud can be used to impose an additional constrain on the parameter $T_c$ introduced in Sec.~\ref{sec:protons}.
The intensity of the Fe \rom{1} K$\alpha$ line emission from the region of the molecular cloud can be calculated as:
\begin{eqnarray}
F_{6.4\text{keV}}=\frac{M_{cl}}{4\pi D_s^2m_{avg} }\int_{T_c}^{\infty} \df T_i \sigma_{\text{K}\alpha}(T_i)4\pi J_i(T_i)\label{eq-kalpha}
\end{eqnarray}
where $\sigma_{\text{K}\alpha}(T_i)$ is the cross section of the K$\alpha$ line emission \citep{tatischeff2012}.
A lower limit for $T_c$ can be obtained by imposing $F_{6.4\text{keV}} < F_{ex}$.
This gives $T_c > 30$ MeV, which is consistent with the range of possible values of $T_c$ reported in Table \ref{tab-fit}.


\section{Discussion and conclusion}

In this paper we derived constraints on the CR proton and electron spectra in the region of the SNR W28. The gamma-ray emission from the MCs in the region demonstrates that an excess of CR protons is present there. We focussed our study to the north-eastern cloud, which is interacting with the SNR shock.

\citet{vaupre2014} tentatively proposed that the excess of CR protons in the region might also explain the enhanced ionization rate observed from the north-eastern cloud. However, also CR electrons and/or X-ray photons coming from the SNR shock can contribute to the ionization.

Here, we developed a model for the transport of X-ray photons into the cloud and demonstrated that their contribution to the observed ionization rate is negligible. Moreover, even though we could not rule out completely CR electrons as the main ionizing agents, we showed that the most natural explanation for the enhanced ionization rate is explained in terms of interactions of CR protons.

To explain both the gamma-ray emission from the cloud and the enhanced ionization rate, the spectrum of protons in the cloud must extend to particle energies smaller than those constrained by gamma-ray observations ($\gtrsim 1$ GeV). 
However, an extrapolation of the spectrum of an order of magnitude only in particle energy would suffice to explain both high and low energy observations. This makes protons the most plausible dominant ionizing agents inside the cloud.

The minimal scenario that would explain simultaneously high and low energy data would require the presence of CR protons
characterized by a single power law spectrum of slope $\delta_p \sim 2.8$ extending over a very broad energy range, spanning from $T_c \lesssim 100$ MeV up to several tens of TeV: almost 6 orders of magnitude in energy!
In fact, the presence of a spectral break cannot be ruled out below a particle energy of $\approx 1$ GeV, because the ionization rate depends only on the integral of the CR spectrum over particle energy.
The presence of a break would require to modify the value of $T_c$ accordingly, in order to reproduce correctly the observed ionization rate, but this would not affect in any significant way the main conclusion of our study.

The work presented in this paper shows how the combination of high and low energy observations of SNR/MC systems can be used as a very powerful tool to gather information on the CR spectrum at specific locations in the Galaxy over an energy range of unprecedented breadth. 
More studies in this direction are desirable, as they will shed light on the process of CR acceleration and escape from SNR shocks \citep{gabici2015}.

\begin{acknowledgement}
The authors would like to thank C. Ceccarelli and T. Montmerle for many years of useful discussions. VHMP would like to thank M. K. Erdim, N. Tsuji, H. Yoneda, N. Cesur, and Prof. S. Safi-Harb for helpful discussions during the DIAS Summer School in High-Energy Astrophysics 2018 and D. Allard and V. Tatischeff for support during my thesis. This project has received funding from the European Union’s Horizon 2020 research and innovation programme under the Marie Skłodowska-Curie grant agreement No 665850, from the Agence Nationale de la Recherche (grant ANR- 17-CE31-0014), and
from the Observatory of Paris (Action F\'ed\'eratrice CTA). GM acknowledges the support received through The Grants ASI/INAF n. 2017-14-H.O and SKA-CTA-INAF 2016.
\end{acknowledgement}

\end{document}